\documentclass[a4paper,twocolumn,amsmath,amssymb,superscriptaddress,longbibliography,aps,pre,nofootinbib]{revtex4-1}

\usepackage{graphicx}
\usepackage{bm}
\usepackage{braket}
\usepackage{dsfont}
\usepackage[usenames,dvipsnames]{xcolor}
\usepackage[tight]{subfigure}
\usepackage{verbatim}
\usepackage{units}
\usepackage{natbib}
\usepackage{multirow}
\usepackage{enumitem}
\usepackage{mathrsfs}
\usepackage{leftidx}
\usepackage{xspace}

\usepackage[normalem]{ulem}  

\usepackage[a4paper]{hyperref}
\hypersetup{colorlinks=true,linktoc=all,linkcolor=blue,breaklinks=true,citecolor=blue,urlcolor=blue}
\usepackage[english]{babel}
\AtBeginDocument{%
    \newwrite\bibnotes
    \def\bibnotesext{Notes.bib}
    \immediate\openout\bibnotes=\jobname\bibnotesext
    \immediate\write\bibnotes{@CONTROL{REVTEX41Control}}
    \immediate\write\bibnotes{@CONTROL{%
    apsrev41Control,author="08",editor="1",pages="1",title="0",year="1"}}
     \if@filesw
     \immediate\write\@auxout{\string\citation{apsrev41Control}}%
    \fi
}%

\usepackage{hyperref}

\newcommand{\bea}{\begin{eqnarray}}
\newcommand{\eea}{\end{eqnarray}}
\newcommand{\beq}{\begin{equation}}
\newcommand{\eeq}{\end{equation}}

\newcommand{\rmh}{\text{h}}
\newcommand{\rmr}{\text{r}}
\newcommand{\rmg}{\text{g}}
\newcommand{\rmc}{\text{c}}
\newcommand{\rmC}{\text{C}}
\newcommand{\rmS}{\text{S}}

\newcommand{\rmR}{\text{R}}
\newcommand{\rmG}{\text{G}}

\newcommand{\kB}{k_\text{B}}
\newcommand{\kBT}{k_\text{B}T}

\DeclareMathOperator{\Tr}{Tr}

\begin{document}

\title{Hybrid Thermal Machines: Generalized Thermodynamic Resources for Multitasking}

\author{Gonzalo Manzano}
\affiliation{International Centre for Theoretical Physics, Strada Costiera 11, 34151 Trieste, Italy.}
\affiliation{Institute for Quantum Optics and Quantum Information (IQOQI), Austrian Academy of Sciences, Boltzmanngasse 3, 1090 Vienna, Austria.}
	
\author{Rafael S\'anchez}
\affiliation{Departamento de F\'isica Te\'orica de la Materia Condensada, Condensed Matter Physics Center (IFIMAC), and Insituto Nicol\'as Cabrera, Universidad Aut\'onoma de Madrid, 28049 Madrid, Spain\looseness=-1}

\author{Ralph Silva}
\affiliation{Institute for Theoretical Physics, ETH Z\"urich, Wolfgang-Pauli-Str. 27, Z\"urich, Switzerland}

\author{G\'eraldine Haack}
\affiliation{D\'epartement de Physique Appliqu\'ee, Universit\'e de Gen\`eve, 1211 Geneva, Switzerland}

\author{Jonatan B. Brask}
\affiliation{Department of Physics, Technical University of Denmark, 2800 Kongens Lyngby, Denmark}

\author{Nicolas Brunner}
\affiliation{D\'epartement de Physique Appliqu\'ee, Universit\'e de Gen\`eve, 1211 Geneva, Switzerland}

\author{Patrick P. Potts}
\affiliation{Physics Department and NanoLund, Lund University, Box 118, 22100 Lund, Sweden}

	\begin{abstract}
Thermal machines perform useful tasks--such as producing work, cooling, or heating--by exchanging energy, and possibly additional conserved quantities such as particles, with reservoirs. Here we consider thermal machines that perform more than one useful task simultaneously, terming these ``hybrid thermal machines''
We outline their restrictions imposed by the laws of thermodynamics and we quantify their performance in terms of efficiencies. To illustrate their full potential, reservoirs that feature multiple conserved quantities, described by generalized Gibbs ensembles, are considered. A minimal model for a hybrid thermal machine is introduced, featuring three reservoirs and two conserved quantities, e.g., energy and particle number. This model can be readily implemented in a thermoelectric setup based on quantum dots, and hybrid regimes are accessible considering realistic parameters. 
\end{abstract}

	\maketitle	


\section{Introduction}
\label{sec:intro}

A considerable body of work has been devoted to the study of thermal machines at the nanoscale in recent years. Interestingly this research covers a broad range of approaches, see e.g.~\cite{binder:2018, benenti:2017,kosloff:2014,mavroidis:2004}. On the one hand some works discuss simple and abstract models in order to derive the basic principles of thermal machines. On the other hand, proposals for practical implementations have been presented as well as first proof-of-principle experiments. 

From the more fundamental and abstract perspective, a promising avenue is to derive minimal models for thermal machines. Characterizing their performance and limits leads to a deeper understanding of fundamental laws of (quantum) thermodynamics. A class of models of particular interest in this context are autonomous thermal machines, which are powered by purely thermal resources and involve only time-independent Hamiltonians and couplings \cite{palao:2001,linden:2010prl,levy:2012,mitchison:2020}. Minimal models, where the machine consists of only few quantum levels, have been thoroughly investigated \cite{brunner:2012,correa:2014, brunner:2014,mitchison:2015,brask:2015, correa:2015pre, manzano:2019}, and provide direct connections to the second and third laws of thermodynamics \cite{skrzypczyk:2011,silva:2016,clivaz:2019prl}. In addition, the basic working principles of these machines can be directly mapped to more realistic devices, which has led to a number of proposals for implementations \cite{venturelli:2013,mitchison:2016,hofer:2016prb,hofer:2016}, as well as first experiments \cite{SingaporeFridge}.

\begin{figure}[b]
\includegraphics[width=0.9\linewidth]{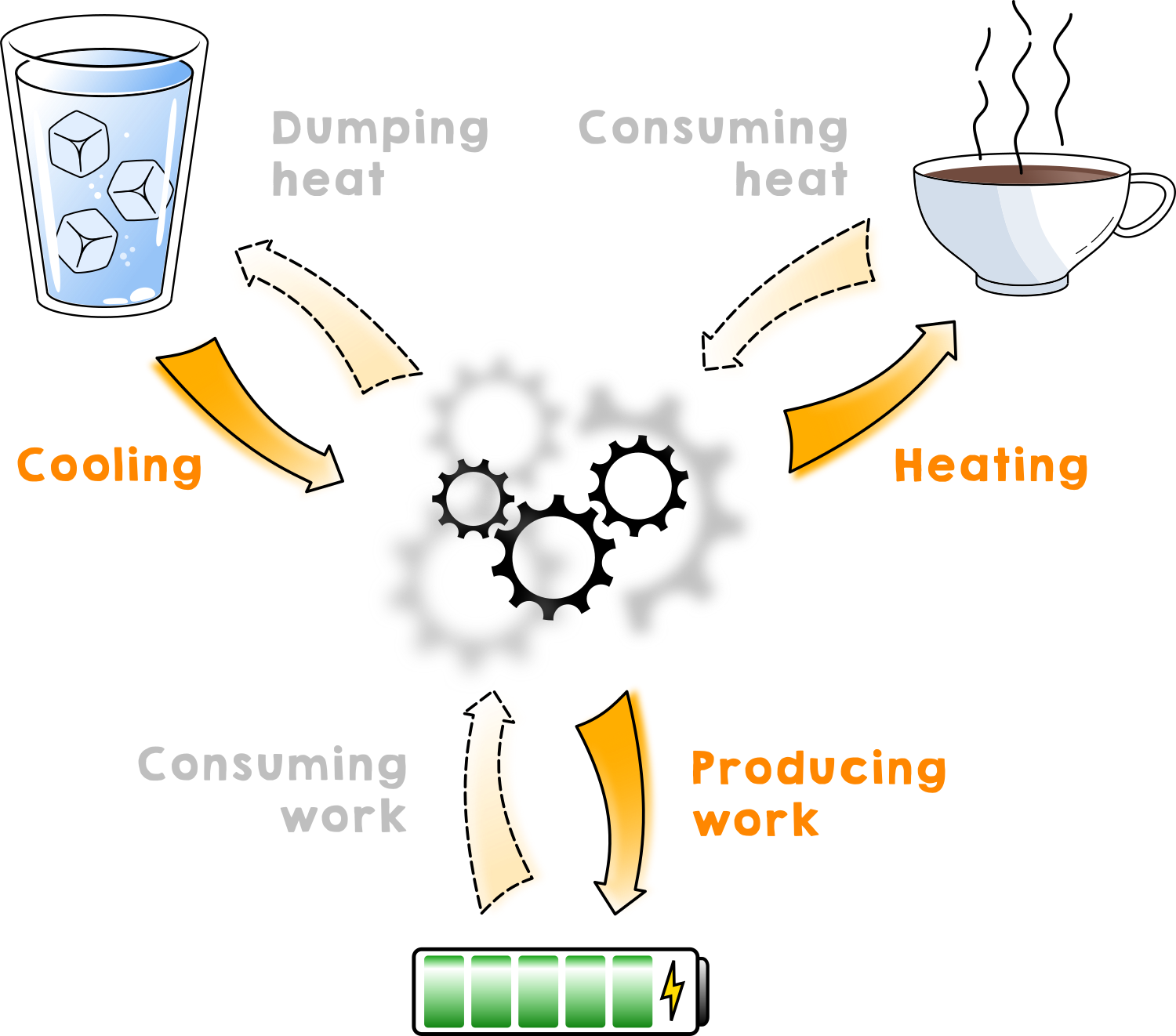}
\caption{A hybrid thermal machine performs multiple useful tasks simultaneously. These tasks include the production of work (engine), cooling of a cold reservoir (refrigerator), and heating of a hot reservoir (heat pump). Multiple energetic resources may be consumed (e.g., heat from a hot reservoir and work). In addition, conserved quantities other than energy might be exchanged, such as particles. }\label{fig:hybridmachine}
\end{figure}

So far, these ideas have been mainly investigated in the scenario where the machine under investigation performs a single task, such as refrigeration, heat pumping, or the production of work. Here we are interested in \textit{hybrid thermal machines}, which are devices that perform more than one useful task simultaneously as illustrated in Fig.~\ref{fig:hybridmachine}. An example of a hybrid regime of operation has been previously investigated in a thermoelectric device~\cite{entin:2015}. There is however no general understanding of hybrid machines, and characterizing their thermodynamic performance is an open challenge. 

Here we follow a general and abstract approach which allows us to characterize general properties of hybrid machines and quantify their efficiency. We explore hybrid thermal machines that, in addition to exchanging energy with their environment, also exchange additional conserved quantities such as particles. The reservoirs acting as resources for the machine may then be described by generalized Gibbs ensembles~\cite{Guryanova2016,YungerHalpern2016}. This approach is well motivated from several perspectives. From a practical point of view, many realistic thermal machines involve the exchange of energy and particles. This naturally leads to a system with several conserved quantities. Typical examples are provided by mesoscopic thermoelectric conductors~\cite{giazotto:2006,sothmann:2015,benenti:2017}. From the more fundamental side, the use of generalized Gibbs ensembles as resources for thermal machines opens novel possibilities and perspectives. Notably, one conserved quantity may now be traded for another, and fundamental limits on these processes are captured by a generalized formulation of the second law~\cite{Guryanova2016,YungerHalpern2016}.

We discuss the implications of our results with the help of a minimal model for a hybrid thermal machine powered by generalized Gibbs ensembles. To this end, we consider a three-terminal device where both energy and one additional conserved quantity (e.g., particles) may be exchanged. We characterize all possible regimes of operation. Beyond the standard thermodynamic tasks (refrigerator, heat pump, and heat engine), we identify a number of hybrid regimes, where two different thermodynamic tasks are performed simultaneously (for example, cooling and power production). Notably, these hybrid regimes exploit both conserved quantities.

In turn, we illustrate the practical relevance of hybrid thermal machines by analyzing an implementation of our minimal model based on two capacitively coupled quantum dots~\cite{molenkamp:1995,chan:2002,hubel:2007,hubel:2008,mcclure:2007,bischoff:2015,hartmann:2015,thierschmann:2015,keller:2016}, in the parameter regime corresponding to recent experiments~\cite{thierschmann:2015}. Moreover, heat engines and refrigerators corresponding to implementations of our minimal model have been experimentally demonstrated in electronic systems, using as a third reservoir the phononic environment~\cite{entin:2010,krause:2011,jiang:2017,dorsch:2020}, or another fermionic reservoir ~\cite{edwards:1993,prance:2009,sanchez:2011,jordan:2013,thierschmann:2015,sothmann:2012,roche:2015,jaliel:2019}.
We note that our approach can be readily extended to other thermodynamic configurations with an arbitrary number of conserved quantities, such as hydrodynamic systems subjected to rotations or translations~\cite{horowitz:2016}, information erasure protocols based on spin-angular momentum~\cite{vaccaro:2011, vaccaro:2013, croucher:2017}, or even quantum squeezed thermal reservoirs~\cite{manzano:2018b}. 

The rest of this paper is structured as follows. In Sec.~\ref{sec:gibbs}, we recapitulate the concept of generalized Gibbs ensembles and provide the first and second laws of thermodynamics. We then define an efficiency for hybrid thermal machines in Sec.~\ref{sec:efficiency}, introducing a reference temperature to quantify the usefulness of different thermodynamic tasks. In Sec.~\ref{sec:hybrid}, we introduce a minimal model for hybrid thermal machines and provide a detailed discussion of the different regimes of operation that are achievable. A physical implementation based on quantum dots is used to illustrate our results in Sec.~\ref{sec:setup}. We discuss the implications of using different reference temperatures for quantifying the usefulness of thermodynamic tasks in Sec.~\ref{sec:reference}, and conclude in Sec.~\ref{sec:conclusions}.

\section{Thermal machines of multiple conserved quantities}
\label{sec:gibbs}

The thermal machines we consider consist of a system which is in contact with multiple reservoirs that are in local equilibrium. To introduce reservoirs with multiple conserved quantities, we review the notion of the equilibrium state. If energy is the only conserved quantity, then the equilibrium state of a quantum system is of the standard Gibbs form, $\tau = e^{-\beta H}/\mathcal{Z}$, where $\beta = 1/(\kBT)$ is the inverse temperature and $\mathcal{Z} = \Tr[e^{-\beta H}]$ the partition function. In the presence of additional conserved quantities $\{A^\alpha\}$, the equilibrium state takes on the form~\cite{Guryanova2016,YungerHalpern2016}
\begin{equation}\label{eq:GGE}
    \tau 
        = e^{- \beta \left( H - \sum_\alpha \mu^\alpha A^\alpha \right) } / \mathcal{Z},
\end{equation}
where the partition function is now $\mathcal{Z} = \Tr[ e^{- \beta \left( H - \sum_\alpha \mu^\alpha A^\alpha \right) }]$. The  ``potentials" $\{\mu^\alpha\}$ play a role analogous to the inverse temperature to the various $A^\alpha$. Take a reservoir defined by $\{\beta, \mu^\alpha\}$ and label the physical quantities that flow out of the reservoir
by $\dot{E}$ and $\dot{A}^\alpha$. Then the contribution to the entropy production due to the exchange of these quantities is~\cite{note}
\begin{align}\label{eq:entropyGGE}
    \dot{S} &= -\beta \left( \dot{E} - \sum_{\alpha} \mu^\alpha \dot{A}^\alpha \right) = - \beta \dot{Q},
\end{align}
where $\dot{Q} = \dot{E} - \sum_\alpha \mu^\alpha \dot{A}^\alpha$ denotes the heat flow from the reservoir.
From the association of the energy flow to the sum of heat and work, $\dot{E} = \dot{Q} - \dot{W}$ we  identify the overall output power as
\begin{align}
    \dot{W} &= -\sum_\alpha \mu^\alpha \dot{A}^\alpha.
\end{align}
Note that positive $\dot{W}$ increases the reservoir energy in contrast to positive $\dot{Q}$. Since the $A^\alpha$ are independently conserved, it is appropriate to split the above into each type of work, $\dot{W}^\alpha = - \mu^\alpha \dot{A}^\alpha$. Note that the $A^\alpha$ may represent entirely different quantities, such as particle number and angular momentum.

All of this concerns a single reservoir. Autonomous thermal machines comprise multiple reservoirs at different temperatures (and potentials) and make use of currents between the various reservoirs to power processes such as refrigeration and the production of work. For this reason, we perform a thermodynamic analysis of several connected reservoirs in the context of generalized Gibbs ensembles. To this end, we consider a collection of reservoirs (labelled by subscripts $i$), each of which has equilibrium temperature $\{\beta_i\}$ and potentials $\{\mu_i^\alpha\}$. A system exchanges energy $E_i$ and the other conserved quantities $A^\alpha_i$ with each reservoir. We note that in the presence of multiple reservoirs, it may happen that the same physical quantity is independently conserved in different regions. This may happen, e.g., in a system of capacitively coupled conductors, each of them having charge conserved independently~\cite{polettini:2016,rao:2018}. 

We assume that in the long-time limit, the machine reaches a nonequilibrium steady state, characterized by a set of time-independent currents $\dot{E}_i$ and $\dot{A}_i^\alpha$ flowing from the reservoirs into the system.
These fluxes are not arbitrary, as the laws of thermodynamics impose fundamental limits on their signs and magnitudes, therefore constraining the possible regimes of operation~\cite{benenti:2017}. The conservation of energy $\sum_i \dot{E}_i=0$ can be cast into the first law of thermodynamics
\begin{equation}
    \label{eq:first}
    \sum_\alpha \dot{W^\alpha} = \sum_i \dot{Q}_i,
\end{equation}
where $\dot{W}^\alpha = -\sum_i \mu_i^\alpha \dot{A}_i^\alpha$ represents the total power output of the physical quantity $\alpha$, and $\dot{Q}_i = \dot{E}_i - \sum_\alpha \mu_i^\alpha \dot{A}_i^\alpha$ denotes the average heat current from reservoir $i$.

The second law of thermodynamics imposes the non-negativity of the total entropy production rate~\cite{spohn:2007,kosloff:2014}
\begin{equation} \label{eq:second}
\dot{S}_{\mathrm{tot}} = -\sum_i \beta_i \dot{Q}_i \geq 0.
\end{equation}
Combining Eqs.~\eqref{eq:first} and \eqref{eq:second} the thermodynamic possibilities of the configuration can be determined. For example, simultaneous cooling of all reservoirs, i.e., $\dot {Q}_i \geq 0 ~~\forall\, i$ is forbidden by the second law. However, the laws of thermodynamics do allow multiple useful tasks to be carried out simultaneously. For instance, a machine that is both an engine and a refrigerator may produce work and cool the coldest reservoir at the same time.

\section{How efficient are hybrid thermal machines?}
\label{sec:efficiency}
In thermal machines that perform a single useful task, a single output is obtained such as produced work, or heat extracted from a cold reservoir. Furthermore, a single input can usually be identified, for instance the heat provided by a hot reservoir or consumed work. An efficiency can then be identified by simply dividing the output by the input. In hybrid thermal machines, multiple outputs are generated using multiple inputs. Quantifying and comparing the usefulness of different outputs and inputs then becomes a non-trivial task raising questions such as: Is producing work more or less useful than cooling a cold reservoir? Compared to cooling a cold reservoir, how much more useful is it to cool an even colder reservoir? Is heat extracted from a hot reservoir less precious than consumed work?

These questions may be settled in an appealing way by introducing a reference temperature $T_{\rm r}$. The reference temperature determines when a reservoir should be considered ``hot'' ($T>T_\rmr$), such that extracting heat from it should be seen as wasteful, and when it should be considered ``cold'' ($T<T_\rmr$), such that extracting heat from it should be seen as useful refrigeration. 
 In addition, a reference temperature is crucial to treat heat and work on an equal footing, and compare their usefulness. In particular it allow us to define the free-energy change in reservoir $i$ (with respect to the reference temperature $T_\rmr$) as $\dot{F}_i\equiv-\dot{E}_i-\kBT_\rmr \dot{S}_i$. See App.~\ref{sec:free-currents} for a detailed discussion. With this definition at hand we can cast the second law in terms of a decrease in free energy of all reservoirs
\begin{equation}
\label{eq:secondlawF}
    \dot{F}_{\rm tot} \equiv \sum_i \dot{F}_i=\sum_\alpha \dot{W}_\alpha+\sum_i\dot{Q}_i\left(\frac{T_{\rm r}}{T_i}-1\right)\leq 0,
\end{equation}
The merit of casting the second law in this form is that each term may be associated with a useful or a wasteful process. The useful processes have a positive sign and include the generation of work ($\dot{W}_\alpha>0$), cooling a reservoir colder than $T_{\rm r}$, and heating a reservoir hotter than $T_{\rm r}$. The wasteful processes have a negative sign and include the consumption of work, cooling a hot reservoir, and heating a cold reservoir (see also Fig.~\ref{fig:hybridmachine}). The reference temperature characterizes the usefulness of each heat current $\dot{Q}_i$ with respect to work generation processes through the thermodynamic factors $T_\mathrm{r}/T_i -1$.

Based on Eq.~\eqref{eq:secondlawF}, we introduce an efficiency by dividing all the useful terms (the outputs) by all the wasteful terms (the inputs)
\begin{equation}
\label{eq:efficiency}
    \eta = \frac{\sum^+_{\alpha}\dot{W}^\alpha+\sum_{i}^+\dot{Q}_{i}\left(\frac{T_{\rm r}}{T_{i}}-1\right) }{-\sum_\alpha^-\dot{W}^{\alpha}-\sum_{i}^-\dot{Q}_{i}\left(\frac{T_{\rm r}}{T_{i}}-1\right)}\leq 1,
\end{equation}
where $\sum^\pm_i x_i = \sum_i (x_i\pm |x_i|)/2$ are the sums over the positive and negative terms respectively. The efficiency may reach unity at points of reversibility, where no entropy is generated. These points are generalizations of the Carnot point in heat engines, where heat is converted into work at maximum efficiency but infinitely slowly. While $\eta$ quantifies the efficiency with which all the outputs are generated by all the inputs, it can be written as a sum of efficiencies that characterize a single useful task each, by only keeping a single term in the numerator. Note that the choice of $T_\rmr$ has an impact on the values of both the efficiency $\eta$, as well as on the components associated to single tasks.

While work and heat from a hot reservoir are usual inputs for thermal machines, it is rather unconventional to treat the heat dumped into a cold reservoir as an input, as we do here. A pictorial example may serve to illustrate the different choices of inputs: In a combustion engine, the hot reservoir is created by the combustion of fuel, an exothermic reaction that produces heat. This heat is then a natural choice for the input of the engine.
One may however consider an engine that uses an endothermic reaction that absorbs heat in order to create a cold reservoir. In this case, the absorbed heat is a natural choice for the resource that is used. One may speculate that the abundance of combustion reactions contributed to the canonical choice for the input being heat provided by a hot reservoir.

We remark that the notion of efficiency introduced here differs from previous proposals based on both (a) the simple ratios between output and input energetic currents, and (b) the split of entropy production into positive and negative contributions corresponding to input and output thermodynamic fluxes. On the one hand, we stress that the approach we propose here has a clean connection with the second law of thermodynamics that proposals of type (a) lack, in particular regarding efficiency bounds and the identification of reversible points. Indeed, it has been recently shown in Ref.~\cite{Hajiloo2020} that such efficiencies may lead to divergences in some multiterminal configurations. 
On the other hand, we notice that while approaches of type (b) have been successfully employed in the linear response regime for two-terminal setups (see e.g. Ref.~\cite{Proesmans2016}), an extension to more general situations is not straightforward, and may require again the introduction of a reference in multiterminal configurations.
What is more, we emphasize that hybrid engines cannot in general be split into individual tasks, so a separation of input and output fluxes is not always meaningful.
In this context, we believe that the definition in Eq.~\eqref{eq:efficiency}, while maintaining a tight link with the second law [in Eq.~\eqref{eq:secondlawF}], has a number of advantages in the context of hybrid thermal machines. For example it allows to easily identify the relevant regimes of operation, to split the efficiency for characterizing different tasks that may be performed simultaneously, and can be compared with the efficiency of traditional heat engines, pumps and refrigerators, as we discuss in detail in Secs. \ref{sec:hybrid} and \ref{sec:reference}.

For the inequality in Eq.~\eqref{eq:secondlawF} to be satisfied, at least one term has to be negative, implying at least one wasteful process. Equation \eqref{eq:secondlawF} thus provides an upper limit on the number of useful tasks that can be carried out simultaneously. For a system with $\mathcal{N}_{\rm res}$ reservoirs with $T_i\neq T_\rmr$, and $\mathcal{N}_{\rm qua}$ extra conserved quantities other than energy, the maximal number of parallel tasks is
\begin{equation}
    \label{eq:simult}
    \mathcal{N}_{\rm tasks}
    = \mathcal{N}_{\rm res} + \mathcal{N}_{\rm qua} -1.
\end{equation}
Note that $\mathcal{N}_{\rm res}$ is not equal to the total number of reservoirs if one or more reservoirs are at the reference temperature.

We stress that our results are only based on the assumption of a steady state, and the first and second law of thermodynamics. No assumption on the coupling strength between system and reservoir is made.

\section{A minimal hybrid machine} \label{sec:hybrid}

\begin{figure}[t]
\includegraphics[width=0.9\linewidth,clip] {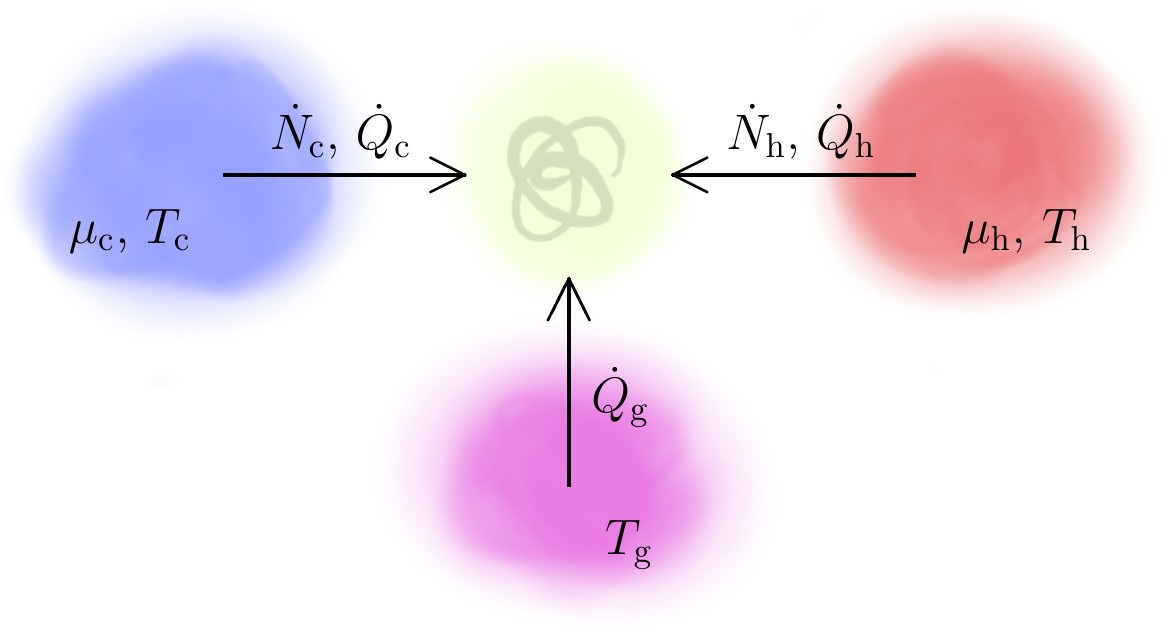}
\caption{\label{fig:scheme} Minimal hybrid machine. A system (represented by the yellow shadow) is coupled to a cold and a hot reservoir (subscripts c and h), with which it exchanges particles and heat, and to a third one with which it only exchanges heat (subscript g). This machine may perform multiple useful tasks simultaneously. For instance, it may cool the cold reservoir and produce work by moving particles against a chemical potential bias $\mu_\rmc-\mu_\rmh$. 
}
\end{figure}

To illustrate these concepts in action, we pick what is arguably the simplest possible configuration that allows us to do so: a three-terminal device, featuring a single additional conserved quantity (other than energy), as sketched in Fig,~\ref{fig:scheme}. For concreteness, we consider the additional conserved quantity to be particle number and we have a thermoelectric device in mind. The results in this section are however completely general. For simplicity, we only allow two of the reservoirs to exchange particles, the cold (subscript c) and the hot (subscript h) reservoir. The third reservoir is denoted by a subscript g for gate. 
We denote the particle flow out of reservoir $i = \rmc,\rmh,\rmg$ by $\dot{N}_i$. The produced power may then be written as $\dot{W}=-\Delta\mu\dot{N}_\rmc$ in the presence of a chemical potential difference $\Delta\mu\equiv\mu_\rmc - \mu_\rmh$, where we have used the conservation of particles, $\dot{N}_\rmc=-\dot{N}_\rmh$. If not explicitly stated otherwise, we will consider the temperatures to fulfill
\begin{equation}
    \label{eq:temps}
    T_\rmc< T_\rmr = T_\rmg< T_\rmh,
\end{equation}
choosing the temperature of the gate as the reference temperature. With this choice, the useful processes correspond to the production of work (heat engine), the cooling of the cold reservoir (refrigerator), and the heating of the hot reservoir (heat pump). Exchanging heat with the gate is considered neither useful nor wasteful. According to Eq.~\eqref{eq:simult}, at most two useful tasks can then be carried out at the same time as we have two reservoirs at $T_i\neq T_\rmr$ and one conserved quantity other than energy. As shown in Sec.~\ref{sec:setup}, such a hybrid regime of operation is indeed realizable in a quantum-dot architecture.

\subsection{Two-terminal device}
\label{sec:twoterm}

Before turning to a description of the three-terminal device, it is useful to consider the regimes of operation that can be obtained by only using two reservoirs, the cold and the hot one. In such a two-terminal setup, the chemical potentials drive a particle flow from the reservoir with higher to the one with lower chemical potential. At the same time, the temperatures induce a heat flow from hot to cold. One of these flows may be utilized to drive the other against its natural direction. 

This may result in a heat engine, when heat flows from hot to cold and induces a particle current against the chemical potential (i.e., $\dot{Q}_\rmh> 0$, $\dot{Q}_\rmc<0$, and $\dot{W}>0$). In a conductor, this implies a thermoelectric effect. The efficiency of the work production is described by Eq.~\eqref{eq:efficiency} which reduces to
\begin{equation}
    \label{eq:etahe2t}
    \eta_{\rm E} = \frac{\dot{W}}{\dot{Q}_\rmh\left(1-\frac{T_\rmr}{T_\rmh}\right)+|\dot{Q}_\rmc|\left(\frac{T_\rmr}{T_\rmc}-1\right)},
\end{equation}
where the subscript stands for \textit{engine}.
The fact that this quantity depends on the reference temperature, even though no reservoir connected to the system is at this temperature, reflects our choice for quantifying heat as a resource. When $T_\rmr \rightarrow T_\rmc$, we consider the heat from the hot reservoir to be the sole resource and Eq.~\eqref{eq:etahe2t} reduces to 
\begin{equation}
    \label{eq:etahe2t2}
    \eta = \frac{1}{\eta_\rmC}\frac{\dot{W}}{\dot{Q}_\rmh},\hspace{1.5cm}\eta_\rmC=1-\frac{T_\rmc}{T_\rmh},
\end{equation}
which is the standard efficiency divided by the Carnot efficiency. When $T_\rmr\rightarrow T_\rmh$, we consider the hot temperature to be abundant and the resource is instead provided by the ability of the cold reservoir to receive heat. In this case, the efficiency reduces to 
\begin{equation}
    \label{eq:etahe2t3}
    \eta = \epsilon_\rmR\frac{\dot{W}}{\dot{Q}_\rmc},\hspace{1.5cm}\epsilon_\rmR=\frac{T_\rmc}{T_\rmh-T_\rmc},
\end{equation}
where $\epsilon_\rmR$ denotes the coefficient of performance for cooling reversibly. In this case, the heat engine may be viewed as a refrigerator operated in reverse. For any choice of $T_\rmr$ in between the hot and the cold temperature, both extracting heat from the hot reservoir and dumping heat into the cold reservoir are considered resources, weighed by prefactors depending on the reference temperature, cf.~Eq.~\eqref{eq:secondlawF}. For a reference temperature $T_\rmr > T_\rmh$, extracting heat from the hot reservoir is seen as useful refrigeration, while for $T_\rmr<T_\rmc$, heating the cold reservoir is seen as useful heat pumping. In these cases, the efficiency is no longer given by Eq.~\eqref{eq:etahe2t}.

Similarly to the production of work, one may use the particle current (i.e., work) to invert the natural tendency of heat to flow out of the hot reservoir, such that the device acts as a heat pump. When both reservoirs are being heated  (i.e., $\dot{Q}_\rmh<0$, $\dot{Q}_\rmc<0$, and $\dot{W}<0$), the efficiency of this process is given by
\begin{equation}
    \label{eq:etap3t}
    \eta_{\rm P} = \frac{|\dot{Q}_\rmh|\left(1-\frac{T_\rmr}{T_\rmh}\right)}{|\dot{Q}_\rmc|\left(\frac{T_\rmr}{T_\rmc}-1\right)+|\dot{W}|},
\end{equation}
where the subscript stands for \textit{pump}. Here, both the consumed work, as well as the heat absorbed by the cold bath are considered resources.
It is again illustrative to consider the limits where the reference temperature coincides with one of the reservoir temperatures. In the limit $T_\rmr\rightarrow T_\rmc$, the efficiency reduces to
\begin{equation}
    \label{eq:etap2t}
    \eta = \eta_\rmC\frac{|\dot{Q}_\rmh|}{\dot{W}}.
\end{equation}
We note that $1/\eta_\rmC$ coincides with the coefficient of performance for heating reversibly. In the limit $T_\rmr\rightarrow T_\rmh$, heating the hot reservoir is no longer considered useful and the efficiency vanishes.

Finally, work can be used to cool down the cold reservoir and heat up the hot reservoir (i.e., $\dot{Q}_\rmh<0$, $\dot{Q}_\rmc>0$, and $\dot{W}<0$), such that the device simultaneously acts as a refrigerator and a heat pump. This is our first example of a hybrid machine, which performs more than one useful task at once. Note however that the second law of thermodynamics prevents operating the device as a refrigerator \textit{without} heating the hot bath at the same time. The efficiency in this regime reads
\begin{equation}
    \label{eq:etafrp2t}
    \eta_{\rm RP} = \frac{\dot{Q}_\rmc\left(\frac{T_\rmr}{T_\rmc}-1\right)+|\dot{Q}_\rmh|\left(1-\frac{T_\rmr}{T_\rmh}\right)}{\dot{W}} = \eta_{\rm RP}^\rmR+\eta_{\rm RP}^{\rm P},
\end{equation}
where the subscript stands for \textit{refrigerator - pump}. In the last equality, we wrote the efficiency as a sum of two terms which correspond to cooling (R) and heat pumping (P) respectively. They are obtained by only keeping the first and second term in the numerator respectively.
When $T_\rmr \rightarrow T_\rmh$, the efficiency reduces to
\begin{equation}
    \label{eq:etaf2t}
    \eta = \frac{1}{\epsilon_\rmR}\frac{\dot{Q}_\rmc}{\dot{W}},
\end{equation}
and characterizes the useful task of cooling the cold reservoir. When $T_\rmr \rightarrow T_\rmc$, only heating the hot reservoir is considered a useful task and the efficiency reduces to Eq.~\eqref{eq:etap2t}.

We thus see that in a two-terminal setup, the efficiency describes a hybrid regime only if the reference temperature does not coincide with either $T_\rmc$ or $T_\rmh$, in agreement with Eq.~\eqref{eq:simult}. As discussed below, in a three-terminal device, hybrid regimes may be implemented that perform more than one useful task no matter the choice of the reference temperature.

\subsection{Three-terminal device} \label{sec:threeterm}
We now consider making use of all three reservoirs, allowing the gate to exchange heat with the system.
This allows for implementing additional regimes of operation.

\subsubsection{Heat engine}
We define the heat-engine operation regime, as the regime where work is produced, i.e., $\dot{W}>0$, but no other useful task is performed, i.e., $\dot{Q}_\rmc<0$ and $Q_\rmh >0$. From Eq.~\eqref{eq:efficiency}, we find that $\dot{Q}_\rmg$ drops out of the efficiency (because $T_\rmr$ = $T_\rmg$) such that we recover the efficiency for a two-terminal heat engine given in Eq.~\eqref{eq:etahe2t}. Nevertheless, $\dot{Q}_\rmg$ still affects the energy flows, as becomes evident from the first law given in Eq.~\eqref{eq:first}. 

Note that our definition of the efficiency differs from approaches that consider the ratio of generated power and injected heat currents~\cite{mazza:2014,whitney_quantum_2016}.

\subsubsection{Refrigerator}
The refrigerator regime is obtained when heat flows out of the coldest reservoir, i.e., $\dot{Q}_\rmc >0$, but no other useful task is performed, i.e. $\dot{W}<0$ and $\dot{Q}_\rmh>0$. In contrast to what we found for work production, refrigeration genuinely benefits from a third reservoir because it allows for implementing an absorption refrigerator. This type of refrigerator uses the natural tendency of heat to flow from the hot reservoir to the gate, in order to extract heat from the cold reservoir. The efficiency for a refrigerator that makes use of both heat and work reads
\begin{equation}
    \label{eq:etaf3t}
\eta_{\rm R} = \frac{\dot{Q}_\rmc}{\epsilon_{\rm AR}\dot{Q}_\rmh+\frac{T_\rmc}{T_\rmr-T_\rmc}|\dot{W}|},\hspace{.75cm}
    \epsilon_{\rm AR}=\frac{1-\frac{T_\rmr}{T_\rmh}}{\frac{T_\rmr}{T_\rmc}-1}
\end{equation}
This efficiency illustrates the parallel operation of an absorption refrigerator, with reversible coefficient of performance $\epsilon_{\rm AR}$, and a refrigerator using work as a resource which operates between the two coldest temperatures $T_{\rmr}$ and $T_\rmc$, cf. Eq.~\eqref{eq:etahe2t3}.  Note that if the work vanishes (e.g. when $\Delta \mu = 0$), we recover the coefficient of performance for an absorption refrigerator, divided by its value at reversibility.

\subsubsection{Heat pump}
The third regime of operation we consider is a heat pump, defined by heat flowing into the hot reservoir, i.e., $\dot{Q}_\rmh<0$, but no other useful task being performed, i.e., $\dot{W}<0$ and $\dot{Q}_\rmc < 0$. As for the engine, we recover the two-terminal expression for the efficiency, cf.~Eq.~\eqref{eq:etap3t}. In contrast to the refrigerator, we do not recover the standard coefficient of performance for an absorption heat pump \cite{brunner:2012} when the work vanishes. The reason for this is that the standard efficiency is obtained by considering the heat injected from the gate as the resource, corresponding to the choice $T_\rmr =T_\rmc$.

\subsubsection{Hybrid regimes}
As discussed above, the second law forbids the simultaneous operation of a heat engine (E) refrigerator (R), and a heat pump (P) given our choices for temperatures in Eq.~\eqref{eq:temps}. However, any two of these tasks may be carried out simultaneously. Three hybrid regimes are therefore possible
\begin{itemize}
    \item ER: Simultaneous production of work and cooling of the cold reservoir ($\dot{W}>0$, $\dot{Q}_\rmc>0$, $\dot{Q}_\rmh>0$).
    \item EP: Simultaneous production of work and heating of the hot reservoir ($\dot{W}>0$, $\dot{Q}_\rmc<0$, $\dot{Q}_\rmh<0$).
    \item RP: Simultaneous cooling of the cold and heating of the hot reservoir ($\dot{W}<0$, $\dot{Q}_\rmc>0$, $\dot{Q}_\rmh<0$).
\end{itemize}
We note that these regimes are obtained by running the single-task regimes in reverse. For instance, regime ER is obtained by running a heat pump that uses both work and heat as resources in reverse. The efficiencies are thus closely related to the efficiencies of the single-task regimes. We find
\begin{equation}
    \label{eq:etaer}
    \eta_{\rm ER} = \frac{\dot{W}+\dot{Q}_\rmc\left(\frac{T_\rmr}{T_\rmc}-1\right)}{\dot{Q}_\rmh\left(1-\frac{T_\rmr}{T_\rmh}\right)} = \eta_{\rm ER}^{\rm E}+\eta_{\rm ER}^{\rm R},
\end{equation}
and
\begin{equation}
    \label{eq:etaer}
    \eta_{\rm EP} = \frac{\dot{W}+|\dot{Q}_\rmh|\left(1-\frac{T_\rmr}{T_\rmh}\right)}{|\dot{Q}_\rmc|\left(\frac{T_\rmr}{T_\rmc}-1\right)}= \eta_{\rm EP}^{\rm E}+\eta_{\rm EP}^{\rm P},
\end{equation}
where the efficiencies with superscripts account for the contributions of every single task (E, R, or P) to the hybrid efficiencies.
The efficiency for the regime RP takes the same form as in the two-terminal setup and is given in Eq.~\eqref{eq:etafrp2t}.
We note that a regime analogous to our regime ER was identified in a thermoelectric device in Ref.~\cite{entin:2015}.

\subsubsection{Simultaneous cooling, heating, and work production}\label{sec:ERP}
Choosing a reference temperature that differs from all three reservoir temperatures, the maximal number of useful tasks that may be performed simultaneously increases to three, see Eq.~\eqref{eq:simult}. Furthermore, simultaneous refrigeration, heat pumping, and work production is possible by choosing $
T_\rmg$ outside the range $[T_\rmc,T_\rmh]$ while still keeping the reference temperature within this range. The efficiency for this triple hybrid regime reads
\begin{equation}
    \label{eq:eterp}
    \eta_{\rm ERP} = \frac{\dot{W}+\dot{Q}_\rmc\left(\frac{T_\rmr}{T_\rmc}-1\right)+|\dot{Q}_\rmh|\left(1-\frac{T_\rmr}{T_\rmh}\right)}{-\dot{Q}_\rmg\left(\frac{T_\rmr}{T_\rmg}-1\right)}.
\end{equation}
Note that in this regime, the useful tasks are either driven by heat extracted from a very hot gate $T_\rmg>T_\rmh$, or by heat dumped into a very cold gate $T_\rmg < T_\rmc$. The machine does therefore never cool the coldest and heat the hottest reservoir simultaneously.

\section{Coupled quantum dots}
\label{sec:setup}

\subsection{The system}

\begin{figure}[t]
\includegraphics[width=\linewidth,clip] {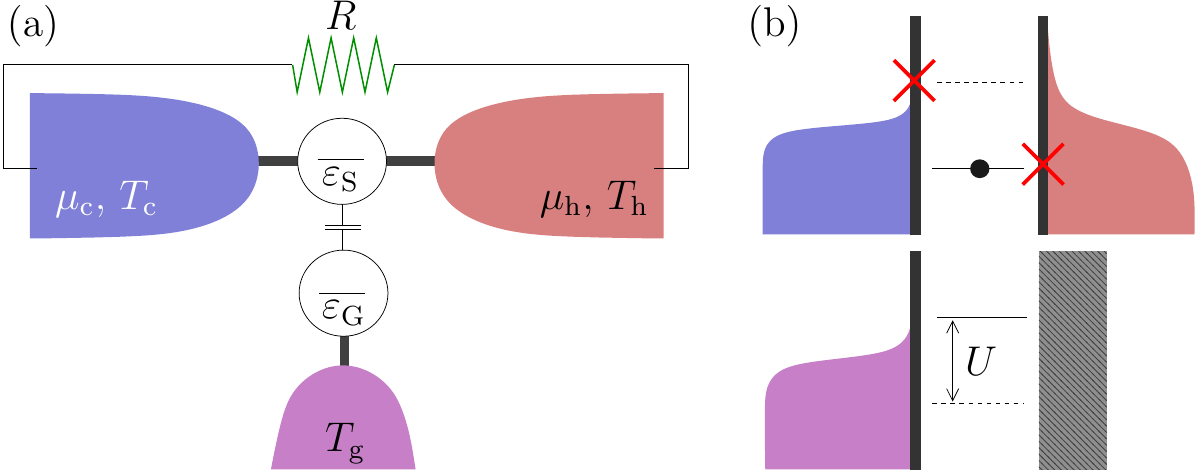}
\caption{\label{fig:system} System based on capacitively coupled quantum dots. (a) The conductor dot (with a level at energy $\varepsilon_\rmS$) is tunnel-coupled to two reservoirs at chemical potential $\mu_l$ and temperature $T_l$ ($l$=c,h). A gate dot (whose level is at $\varepsilon_\rmG$) is coupled to a single reservoir at temperature $T_{\rm g}$. (b) The occupation of one of the dots increases the energy of the other one by $U$. We assume that tunneling to reservoir h with $\varepsilon_\rmS$ (when the gate dot is empty) and to reservoir c at $\varepsilon_\rmS+U$ (when the gate dot is occupied) is suppressed. Work will be produced by the system when a charge current flows powering a load, i.e., against a voltage difference $V=(\mu_\rmc-\mu_\rmh)/e$ established by the load resistance $R$. 
}
\end{figure}

Let us illustrate the previous arguments with a particular example consisting of a three-terminal system of capacitively coupled quantum dots. This configuration, sketched in Fig.~\ref{fig:system}, has been extensively investigated in the context of electrical and thermal transport as a heat engine~\cite{sanchez:2011,thierschmann:2015,dare:2017,walldorf:2017} and a refrigerator~\cite{zhang:2015,sanchez:2017,dare:2019,erdman_absorption_2018,zhang:2020}.
The device has a conductor-gate geometry. The conductor is formed by one of the dots, which is coupled to two reservoirs via tunneling barriers. It hence supports particle, $\dot{N}_j$, and heat currents, $\dot{Q}_j$, in reservoirs $j=\rmc,\rmh$. The other quantum dot, which we call the gate dot, is only coupled to reservoir g. The capacitive coupling between dots emphasizes that the gate injects no particles into the conductor. Energy exchange between the system and the gate is maintained by charge fluctuations in the dots~\cite{sanchez:2011,ruokola:2011} and mediated by the Coulomb repulsion $U$. 

We will focus on the strong Coulomb blockade regime, where due to electron-electron interactions each dot can be occupied by up to one electron. This way, in the weak coupling limit, the dynamics of the system is described by sequential transitions between the four states $(n_\rmS,n_\rmG)$, with a number of electrons $n_\rmS,n_\rmG=0,1$ in the conductor (S) and gate (G) quantum dots. These are described by the transition rates
\beq
\label{eq:rates}
\Gamma_{j n}^+=\Gamma_{j n}\begin{cases} f_j(\varepsilon_\rmS+nU)\hspace{.2cm}{\rm for}\hspace{.2cm} j=\rmc,\rmh,\\
 f_\rmg(\varepsilon_\rmG+nU)\hspace{.2cm}{\rm for}\hspace{.2cm} j=\rmg,
\end{cases}
\eeq
for an electron tunnelling from terminal $j$ into the adjacent dot, when the other dot contains $n$ electrons. The state of the reservoir, with a chemical potential $\mu_j$  and temperature $T_j$, is described by the Fermi distribution $f_j(E)=1/\{1+\exp[(E-\mu_j)/\kBT_j]\}$ (where we set $\mu_\rmg=0$). The tunneling out transitions, $\Gamma_{j n}^-$, are obtained by replacing $f_j(E)$ by $1-f_j(E)$ in Eq.~\eqref{eq:rates}, such that the transitions fulfill local detailed balance. We assume energy-dependent tunneling rates $\Gamma_{jn}$ that explicitly depend on the charge state of the other dot, $n$. For simplicity we will assume a particular case with $\Gamma_{\rmc1}=\Gamma_{\rmh0}=0$, cf.~Fig.~\ref{fig:system}~(b), where the charge and heat currents are maximally correlated~\cite{sanchez_detection_2012,*sanchez_erratum_2013} such that an electron cannot be transferred between c and h without exchanging the energy $U$ with g. All other transitions are assumed to occur with the same rate, $\Gamma$. 
This very particular tunneling configuration is typically not present in single quantum dot systems. However, it can be obtained by using energy filters provided by, e.g., superconductors or additional quantum dots, cf. Appendix~\ref{sec:exp_lim}. 

The occupation of the system, $P_{n_\rmS n_\rmG}$, is well described by a set of rate equations~\cite{beenakker:1991}. In our case, they read
\begin{align}
\dot{P}_{00}&=\Gamma_{\rmc0}^{-}P_{10}+\Gamma_{\rmg0}^-P_{01}-\left(\Gamma_{\rmc0}^++\Gamma_{\rmg0}^+\right)P_{00},\nonumber\\
\dot{P}_{10}&=\Gamma_{\rmc0}^{+}P_{00}+\Gamma_{\rmg1}^-P_{11}-\left(\Gamma_{\rmc0}^-+\Gamma_{\rmg1}^+\right)P_{10},\\
\dot{P}_{01}&=\Gamma_{\rmh1}^{-}P_{11}+\Gamma_{\rmg0}^+P_{00}-\left(\Gamma_{\rmh1}^++\Gamma_{\rmg0}^-\right)P_{01},\nonumber
\end{align}
with the normalization condition $P_{00}+P_{10}+P_{01}+P_{11}=1$.
From the stationary solution, obtained by solving the system $\dot{P}_{n_\rmS n_\rmG}=0$, we obtain the steady-state currents. For the chosen configuration of tunneling rates, all currents are tightly coupled and proportional to the particle current~\cite{sanchez:2011}
\beq \label{eq:ecurr}
\dot{N}_\rmc = \Gamma_{\rmc0}^+P_{00}-\Gamma_{\rmc0}^-P_{10},
\eeq
such that the heat currents read
\begin{align} \label{eq:tight}
\dot{Q}_\rmc &= (\varepsilon_\rmS - \mu_\rmc) \dot{N}_\rmc ,\nonumber\\ 
\dot{Q}_\rmh &= -(\varepsilon_\rmS + U - \mu_\rmh)\dot{N}_\rmc,  \\ 
\dot{Q}_\rmg &= U \dot{N}_\rmc,   \nonumber
\end{align}
with
\begin{align} 
\label{eq:ecurrexpr}
\dot{N}_\rmc &= A^{-1}\Gamma_{\rmc0}\Gamma_{\rmh1}\Gamma_{\rmg0}\Gamma_{\rmg1}\exp\left(\frac{\varepsilon_\rmG}{\kBT_\rmg}\right) \\
&\times\left[\exp\left(\frac{\varepsilon_\rmS+U-\mu_\rmh}{\kBT_\rmh}\right)-\exp\left(\frac{\varepsilon_\rmS-\mu_\rmc}{\kBT_\rmc}+\frac{U}{\kBT_\rmg}\right)\right]\nonumber\\
&\times f_\rmc(\varepsilon_\rmS)f_\rmh(\varepsilon_\rmS+U)f_\rmg(\varepsilon_\rmG)f_\rmg(\varepsilon_\rmG+U), \nonumber
\end{align}
and with the prefactor $A^{-1}>0$ given by the normalization of the density matrix elements. The power is given by $\dot{W} = -\Delta\mu\dot{N}_\rmc$. Negative $\dot{W}$ is hence interpreted as dissipated Joule heat, while $\dot{W}>0$ means power is generated by the non-equilibrium situation induced by the temperature gradients in the reservoirs. From the expression of $\dot{N}_{\rm c}$, we find that all fluxes vanish at the reversibility point
\beq
\label{eq:revers}
\frac{\mu_\rmh}{T_\rmh}-\frac{\mu_\rmc}{T_\rmc}=
\varepsilon_\rmS\left(\frac{1}{T_\rmc}-\frac{1}{T_\rmh}\right)+U\left(\frac{1}{T_\rmg}-\frac{1}{T_\rmh}\right),
\eeq 
where also $\dot{S}_{\rm tot}=0$.
This is a recurrent feature of heat engines operating in nonequilibrium steady-states due to the tight coupling between the currents~\cite{humphrey:02,odwyer:2005,esposito_thermoelectric_2009, cleuren:12}, which makes all currents vanish simultaneously despite the non-equilibrium situation.

\begin{figure*}[t!]
\includegraphics[width=0.8 \linewidth,clip] {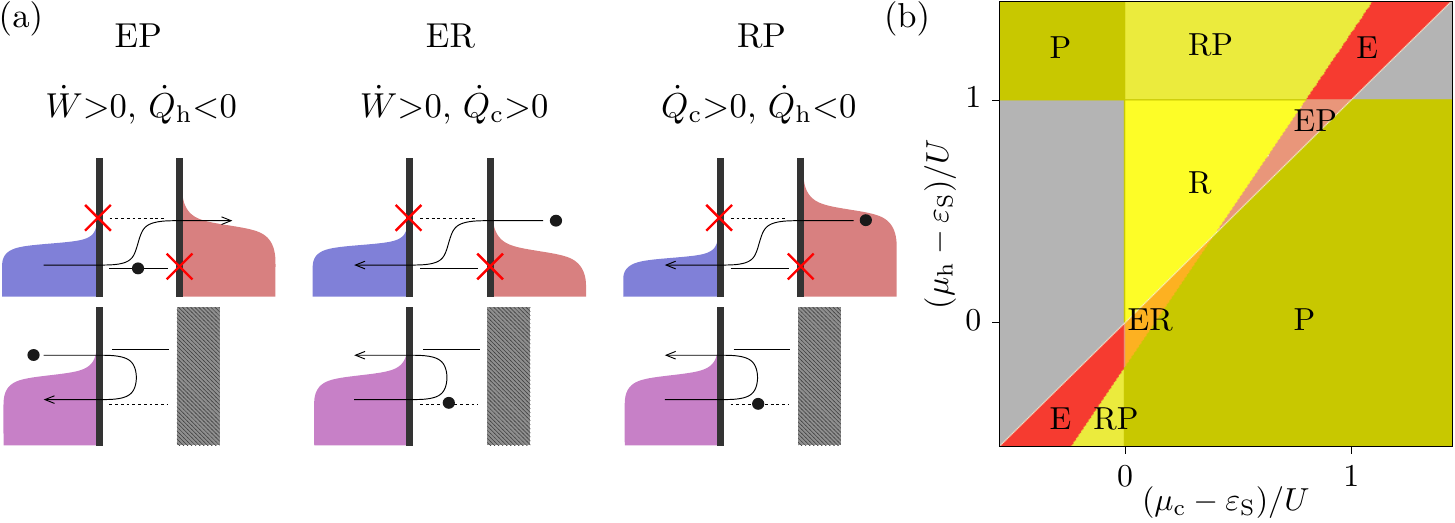}
\caption{\label{fig:scheme_regime} (a) Scheme of the different possible processes leading to the hybrid regimes in our system, EP, ER and RP. Electrons emitted from above the chemical potential of a reservoir $j$ carry positive heat, contributing to cooling it, $\dot{Q}_j>0$. The sign changes for electron tunneling in the opposite direction or to states below the chemical potential. 
Note that the tunneling asymmetry prevents the occurrence of a purely two-terminal RP process through the lowest energy level of the system quantum dot.
(b) Map of the system heat currents and generated power as a function of the chemical potentials.
Power is generated in reddish areas and dissipated in the yellowish ones. In grey regions, no useful task is performed. 
Here, $\varepsilon_\rmS=0$, $\varepsilon_\rmG=-50$, $\kB T_{\rm c}=20$, $\kB T_{\rm r}=25$, $\kB T_{\rm h}=30$, $U=90$ (energies are in meV), with tunneling rates as sketched in (a). 
}
\end{figure*}

\subsection{Hybrid regimes of operation}

The versatility of this setup is manifested when exploring the different regimes of operation discussed above. In particular, we consider the possibility of reproducing the hybrid configurations
EP, ER and RP. In the coupled quantum dot system, the processes giving rise to such operations are sketched in Fig.~\ref{fig:scheme_regime}(a). For temperatures fulfilling Eq.~\eqref{eq:temps}, all possible operations (hybrid or not) can be found in our device by tuning the chemical potentials of the conductor, as shown in Fig.~\ref{fig:scheme_regime}(b),
using typical parameters in state-of-the-art experiments~\cite{thierschmann:2015}.
The parameter configurations where the hybrid regimes occur are delimited by the reversibility condition Eq.~\eqref{eq:revers} and by $\mu_\rmc=\mu_\rmh$, $\mu_\rmc=\varepsilon_\rmS$ and $\mu_\rmh=\varepsilon_\rmS+U$ which define the points at which the different currents change sign according to Eqs.~\eqref{eq:tight}. 

\begin{figure*}[t]
\includegraphics[width=1.0\linewidth] {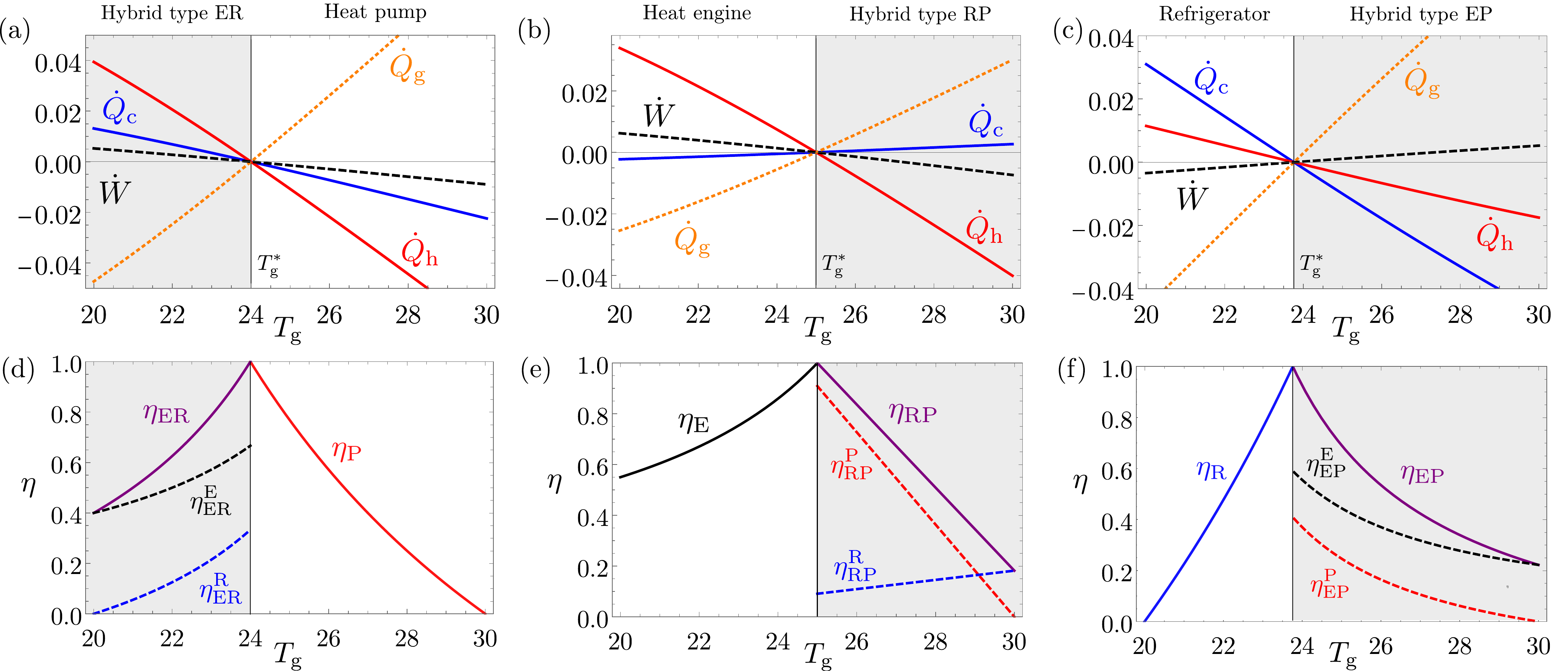}
\caption{\label{fig:currents} (a)-(c) Heat currents and output power  and (d)-(f) hybrid efficiencies together with their contributions, as a function of the gate terminal temperature $T_\rmg$ in the interval $[T_\rmc, T_\rmh]$ with $\kB T_\rmc = 20$ and $\kB T_\rmh = 30$. Different regimes of operation are separated by the values $T_\rmg^\ast$ and obtained for different regimes of parameters: (a) $U = 90$, $\varepsilon_\rmG = -80$, $\mu_\rmc = 25$ and $\mu_\rmh = 15$. (b) $U = 90$, $\varepsilon_\rmG = -80$, $\mu_\rmc = 12$ and $\mu_\rmh = -10$. (c) $U = 40$, $\varepsilon_\rmG = -50$, $\mu_\rmc = 27$ and $\mu_\rmh = 30$. In all cases $\varepsilon_\rmS = 0$ and $\hbar \Gamma = 0.01$. Energy units in the parameters are given in meV while power and heat currents are plotted in units of $10 \times U \Gamma$.    
}
\end{figure*}

We can compute the total efficiency in each regime of the device by directly applying Eq.~\eqref{eq:efficiency} as discussed in Sec.~\ref{sec:threeterm}. 
In Fig.~\ref{fig:currents} we show the heat currents [top panels (a)-(c)], as well as the corresponding efficiency [bottom panels (d)-(f)]. Each column corresponds to a different hybrid regime. Varying $T_\rmg$ allows for switching between a hybrid regime and the complementary single-task regime by crossing the point of reversibility [cf.~Eq.~\eqref{eq:revers}]
\begin{equation}
\kB T_\rmg^\ast \equiv U \left(\frac{\varepsilon_\rmS + U - \mu_\rmh}{\kB T_\rmh} - \frac{\varepsilon_\rmS - \mu_\rmc}{\kB T_\rmc}\right)^{-1},
\end{equation}
where the efficiency reaches unity.

For the hybrid regimes ER, RP and EP, we include the two separate contributions associated to the two different tasks being performed simultaneously in Figs.~\ref{fig:currents}~(d)-(f). Here we see that the efficiencies depend on how the corresponding task(s) are useful with respect to the reference temperature. For example in Fig.~\ref{fig:currents}~(d) we see that when $T_\rmg \rightarrow T_\rmh$ (right part) heat pumping in the hot reservoir may be considered less and less useful and, consequently, $\eta \rightarrow 0$ in this regime. Analogously in Fig.~\ref{fig:currents}~(f) we see a similar behavior: when $T_\rmg \rightarrow T_\rmc$ refrigeration of the coldest reservoirs is less useful with respect to $T_\rmr$, implying $\eta \rightarrow 0$. We can appreciate a similar behavior for the corresponding component efficiencies in the hybrid regimes ER and EP. In Fig.~\ref{fig:currents}~(e) we can indeed see how the two efficiencies composing the RP regime exchange their roles, since refrigeration becomes more useful than heat pumping when $T_\rmg$ increases sufficiently. Furthermore, we see that the efficiencies of the different (individual) tasks in hybrid regimes tend to develop lower values for the efficiencies than in non-hybrid regimes (considering similar $T_\rmg$). This is as a consequence of the fact that in the hybrid regimes, a single input has to drive two outputs and it differs from what was predicted in  Ref.~\cite{entin:2015} using a different expression for the efficiencies.

\subsection{Heating, cooling, and producing work at the same time}
\label{sec:reverse}

Let us now consider the case where the reference temperature is different from that of any of the three terminals. In particular, we choose $T_\rmc<T_{\rm r}<T_\rmh<T_\rmg$ i.e., the gate is the hottest terminal. This configuration allows us to find regimes for which $\dot{W}>0$ with $\dot{Q}_\rmc>0$ and $\dot{Q}_\rmh<0$.
This means that by coupling to the hot gate, the conductor is able to simultaneously produce work, cool its coldest terminal and pump heat into its hottest one, as discussed in Sec.~\ref{sec:ERP}. All charge and heat currents within the conductor are hence reversed. 

This is a very different picture from what one expects to happen in a two terminal conductor that conserves energy, where one has to chose between an operation that moves electrons against a chemical potential difference (work production) or extracting heat from the coldest reservoir (cooling), as discussed in Sec.~\ref{sec:twoterm}. Hence, one can argue that coupling to the gate has a related effect to coupling to an autonomous version of Maxwell's demon~\cite{koski:2015}. Of course, we are not challenging any law of thermodynamics here because the system is using heat injected from the hottest terminal (the gate) as a resource.

\begin{figure}[b]
\includegraphics[width=\linewidth,clip] {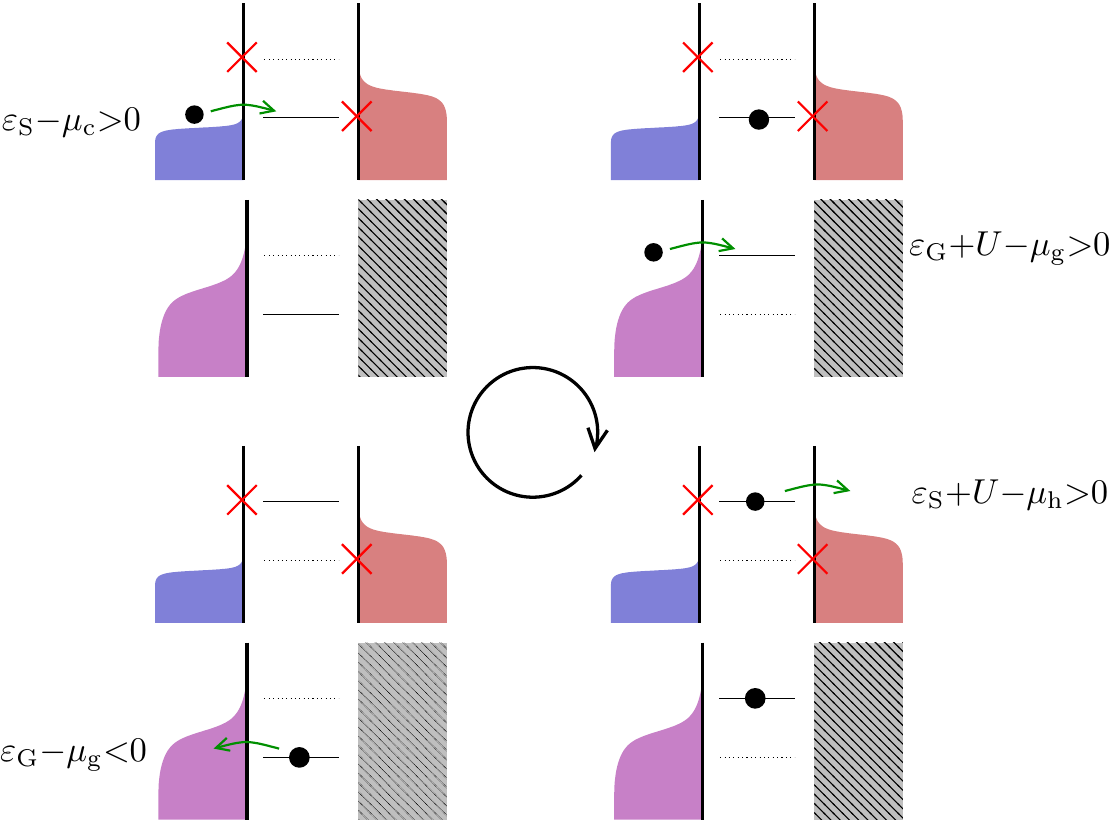}
\caption{\label{fig:cycle} Tunneling sequence that achieves the reversal of heat and charge currents in the two-terminal conductor. An energy $\pm U$ due to interdot Coulomb interaction is exchanged between the conductor and gate after the completion of the cycle clockwise or anticlockwise.
}
\end{figure}

In our system, with the tunneling rates as fixed above, this will be the case when $\Delta\mu<0$. We can understand this in terms of inelastic processes as illustrated in the cycle in Fig.~\ref{fig:cycle}: For low chemical potential differences, the gate terminal being very hot induces the electrons in the conductor to absorb energy when going between the cold and the hot terminal. Then, if $\varepsilon_\rmS>\mu_\rmc$, electrons tunnel from the cold terminal into the dot carrying a positive amount of heat and hence cooling it. The electron then tunnels out to the hot terminal at a higher energy $\varepsilon_\rmS+U>\mu_\rmh$ therefore heating the hot terminal. This is not possible if $T_\rmc<T_\rmg<T_\rmh$.

This regime of operation region is obtained when $\Delta\mu<0$, $\varepsilon_\rmS>\mu_\rmc$ and $\varepsilon_\rmS-\mu_\rmc<E^*(\Delta\mu)$, where 
\begin{equation}
E^*(\Delta\mu)=\left[U\left(1-\frac{T_\rmh}{T_\rmg}\right)- \Delta\mu\right]\left(\frac{T_\rmh}{T_\rmc}-1\right)^{-1}
\end{equation}
is obtained from Eq.~\eqref{eq:revers}. Note that if either $U=0$ or $T_\rmh=T_\rmg$, we get $E^*(0)=0$, so the three conditions would then cross at $\mu_\rmc=\mu_\rmh$. This emphasizes the importance of the Coulomb interaction and the requirement that the gate be hotter than the system. The same happens trivially if $T_\rmc=0$, as no cooling can take place.

\begin{figure}[t]
\includegraphics[width= \linewidth,clip] {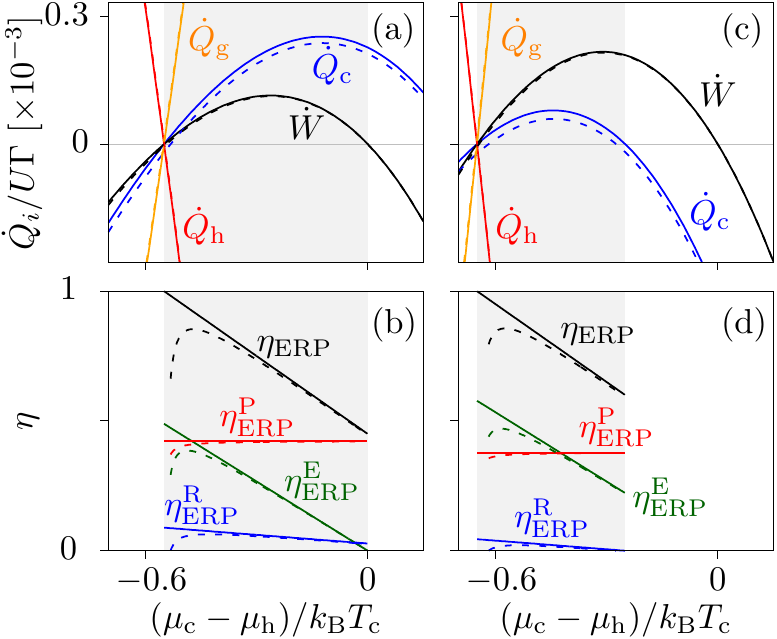}
\caption{\label{fig:currrevref} (a), (c) Currents and (b), (d) efficiencies as functions of the chemical potential difference for simultaneous heating, cooling, and power production (grey shaded area). For (a) and (b), $\varepsilon_{\rmS}=0.25\kBT_\rmc$, for (c) and (d), $\varepsilon_{\rmS}=-0.25\kBT_\rmc$. We use $T_{\rm r}=(T_\rmc{+}T_\rmh)/2$ as the reference temperature. Here, $\kB T_{\rmc}=20$, $\kB T_{\rmh}=25$, $\kB T_{\rmg}=30$, $U=90$ (energies expressed in meV), and $\varepsilon_\rmG=-U/2$. Dashed lines correspond to the case where $\Gamma_{{\rm c}1}=\Gamma_{{\rm h}0}=10^{-3}\Gamma$, such that the tight-coupling condition is lifted.
}
\end{figure}

The introduction of the reference temperature captures this effect nicely: cooling and pumping are well-defined with respect to the reference temperature if it is chosen to be $T_\rmc< T_{\rm r}< T_\rmh$. Then, the efficiency takes the proper processes into account, with three useful operations out of a single resource.
Fig.~\ref{fig:currrevref} shows the different currents and the corresponding efficiencies for $T_{\rm r}=(T_{\rm c}+T_{\rm h})/2$. The hybrid efficiency is described by Eq.~\eqref{eq:eterp}. Note that, for a fixed system configuration, the pumping efficiency, $\eta_{\rm ERP}^{\rm P}$ is independent of the applied voltage due to the tight-coupling condition.

\section{Choosing the reference temperature}
\label{sec:reference}
Throughout the paper we made use of the notion of a reference temperature to characterize the usefulness of different thermodynamic tasks performed by hybrid machines and quantify their performance. The choice of the reference temperature determines the thermodynamic weights assigned to the different processes occurring within the machine. 
As a result, taking a particular reference we may conclude that one task is being performed more efficiently than the other, while other choices of the reference temperature will lead to the opposite conclusion. 
Nevertheless, when considering specific regimes of operation (as in the cases discussed above), there is often a choice for the reference temperature that appears to be natural. Indeed, this choice may vary from one regime of operation to another. In our analysis above, we chose to keep the reference temperature fixed at $T_\mathrm{r} = T_\mathrm{g}$ for all regimes of operation. This explains why our notion of efficiency coincides with the usual notions for certain cases (e.g. absorption refrigerator), but not for others (e.g. absorption heat pump).
We note that while different reference temperatures may result in different quantitative values for characterizing the performance of the machine, the points where reversibility is attained (i.e. $\eta=1$) are always the same.

One may wonder whether a different approach for setting the reference temperature could recover the standard single-task notions of efficiencies from Eq.~\eqref{eq:efficiency}. In the following, we show this to be the case, and discuss the physical meaning.
Let us take a look back at standard machines, performing a single task. Here the reference temperature is typically set to coincide with that of the reservoir where entropy (i.e. heat) is dumped into. In the case of hybrid machines, this is slightly more delicate as there might be several reservoirs acting as entropy sinks. We may then set the reference temperature as that of the coldest entropy sink. 
This choice minimizes the overall free energy losses in the operation of the machine under the constraint that $T_\rmr$ coincides with the temperature of one of the entropy sinks. Indeed, Eq.~\eqref{eq:secondlawF} can be rewritten as $\dot{F}_\mathrm{tot} = - k_B T_\rmr \dot{S}_\mathrm{tot}$, implying that the smaller the value of the reference temperature, the smaller the change in the overall free energy (see App.~\ref{sec:free-currents}).

We now apply this idea to the case of the minimal hybrid machine model introduced above. Instead of being always equal to the intermediate temperature, the reference temperature may now vary. For any regime where the cold reservoir is not being cooled, the reference temperature will now correspond to that of the cold reservoir. That is, in the regimes labeled as E, EP and P, we now set $T_\mathrm{r} = T_\mathrm{c}$ into Eq.~\eqref{eq:efficiency}. In the other regimes ER, RP and R we still have $T_\mathrm{r} = T_\mathrm{g}$. In this way, we recover here all standard efficiencies for single-task machines in the corresponding limiting cases, which allows for a direct comparison with previous works. While this approach works well for the case of our minimal model of a hybrid machine, it would be interesting to further explore its possibilities and limitations in more general configurations.

\section{Conclusions}
\label{sec:conclusions}

We have discussed the operation and performance of thermal machines that perform multiple useful tasks simultaneously and are powered by generalized Gibbs ensembles featuring an arbitrary number of additional conserved quantities. Our results provide a systematic method for quantifying the performance of such hybrid thermal machines, enabling future studies of these devices. In particular we presented a minimal model featuring three reservoirs and two conserved quantities, energy and particle number. This model allows for hybrid regimes of operation, where two useful thermodynamic tasks can be implemented simultaneously. An implementation of such a machine, in particular in hybrid regimes, was discussed for a thermoelectric engine based on capacitively coupled quantum dots, considering realistic parameters.

Our work opens a number of questions. First, it would be interesting to see if the functioning of hybrid machines can be characterized in simple terms, as it is the case for standard autonomous machines, using the notion of virtual qubits~\cite{brunner:2012}. Can a similar notion be devised for machines involving multiple conserved quantities? 
Another question is whether quantum effects (such as coherence and entanglement) can play a role, or even enhance the thermodynamic performance. This is known to be the case for the small autonomous three-qubit refrigerator, where entanglement can boost the cooling power~\cite{brunner:2014}. 
It would also be interesting to consider regimes of operation beyond the steady state regime, which we focus on here. Again, it is known that for small autonomous refrigerators, the transient regimes can feature enhanced cooling \cite{brask:2015,mitchison:2015}. Would it be possible to simultaneously enhance two thermodynamic tasks in a hybrid machine by moving to the transient regime? 

We recall that, given the generality of the approach developed here, our methods could be applied to a variety of physical platforms with an arbitrary number of terminals and additional conserved quantities. We already mentioned some examples in the introduction. Furthermore, there exist also many systems in the solid state, that may be particularly well-suited to test and extend our results. A natural candidate is the spin degree of freedom, as the field of spin-caloritronics is well established experimentally by observations of the spin-Seebeck effect~\cite{uchida:2008,ozaeta:2014,kolenda:2016,kolenda:2017}.
Finally, it would be interesting to consider generalized Gibbs ensembles where the additional conserved quantities do not commute with the Hamiltonian~\cite{Guryanova2016,YungerHalpern2016,Lostaglio2017,manzano:2018b,YungerHalpern2020}. While our results for the average currents studied here are still valid, the possibility of finding imprints of non-commutativity at the level of fluctuations in hybrid machines and the determination of their role remains an open question for future research.

\acknowledgments
G.M. acknowledges funding from the European Union's Horizon 2020 research and innovation programme under the Marie Sk\l{}odowska-Curie grant agreement No 801110 and the Austrian Federal Ministry of Education, Science and Research (BMBWF). R.Sa. acknowledges funding from the Ram\'on y Cajal program RYC-2016-2077, and the Spanish Ministerio de Ciencia e Innovaci\'on via grant No. PID2019-110125GB-I00 and through the ``Mar\'ia de Maeztu'' Programme for Units of Excellence in R{\&}D CEX2018-000805-M. R.S. acknowledges funding from the Swiss National Science Foundation via an Ambizione grant PZ00P2\_185986 and the NCCR QSIT. J.B.B. was supported by the Independent Research Fund Denmark. G.H. acknowledges funding from the Swiss National Science Foundation through the starting grant PRIMA PR00P2\_179748 and the NCCR QSIT (Quantum Science and Technology). P.P.P. acknowledges funding from the European Union's Horizon 2020 research and innovation programme under the Marie Sk\l{}odowska-Curie Grant Agreement No. 796700. 

\appendix

\section{Free-energy currents} \label{sec:free-currents}

In this appendix we  motivate the introduction of the reference temperature by discussing its role in the context of nonequilibrium free energy and its relation to optimal work extraction. Note that here, in order to gain full generality, we also add reference potentials associated to the additional conserved quantities.

Consider the changes in the nonequilibrium free energy of the reservoirs, $\dot{F}_i$ in Eq.~\eqref{eq:secondlawF}. These are related to the optimal work that can be extracted by using a thermal reservoir at the reference temperature $T_\rmr$~\cite{parrondo:2015,skrzypczyk:2014,kammerlander:2016,manzano:2018}. This concept can be further extended to the scenario featuring different conserved quantities~\cite{halpern:2016, halpern:2018,manzano:2018b}. Such a generalized free energy, also called nonequilibrium Massieu potential~\cite{rao:2018}, makes use of a reference reservoir characterized by both a reference temperature $T_\rmr$ and a set of reference potentials $\{\mu_\rmr^{\alpha}\}$ associated to the exchange of additional conserved quantities $\{A^{\alpha}\}$. The optimal work extractable from a system in a generic nonequilibrium state $\rho$ (able to exchange energy and additional quantities with the reference reservoir) is then given by the difference of generalized free energy between $\rho$ and the equilibrium state $\tau$ of the reference reservoir~\cite{halpern:2016, halpern:2018, manzano:2018b}.

We apply this generalization to the changes in the free energy of the different terminals, that are related to the changes in the optimal work that may be extracted from them. The generalized free energy current into
terminal $i$, with respect to a  reference reservoir r then reads:
\begin{align} \label{eq:free}
\dot{F}_i &\equiv -(\dot{E}_i - \sum_\alpha \mu_{\rm r}^{\alpha} \dot{A}_i^{\alpha}) - \kB T_{\rm r} \dot{S}_i \nonumber \\
&= \sum_\alpha (\mu_{\rm r}^{\alpha} - \mu_i^{\alpha}) \dot{A}_i^{\alpha} - \left(1 - \frac{T_{\rm r}}{T_i} \right) \dot{Q}_i,      
\end{align}
where in the second line we used the expression for the energy $\dot{E}_i = \dot{Q}_i + \sum_\alpha \mu_i^{\alpha} \dot{A}_i^{\alpha}$ and the (Clausius) entropy fluxes $\dot{S}_i = - \beta_i \dot{Q}_i$ from reservoir $i$,
as given in Eq.~\eqref{eq:entropyGGE}. We note that when no extra conserved quantities are considered $\dot{A}_i^{\alpha} = 0$ for all $\alpha$, and we recover the standard free energy expression for thermal reservoirs introduced in Sec.~\ref{sec:efficiency}.

\label{sec:exp_lim}
\begin{figure*}[bth]
\includegraphics[width=0.85\linewidth]{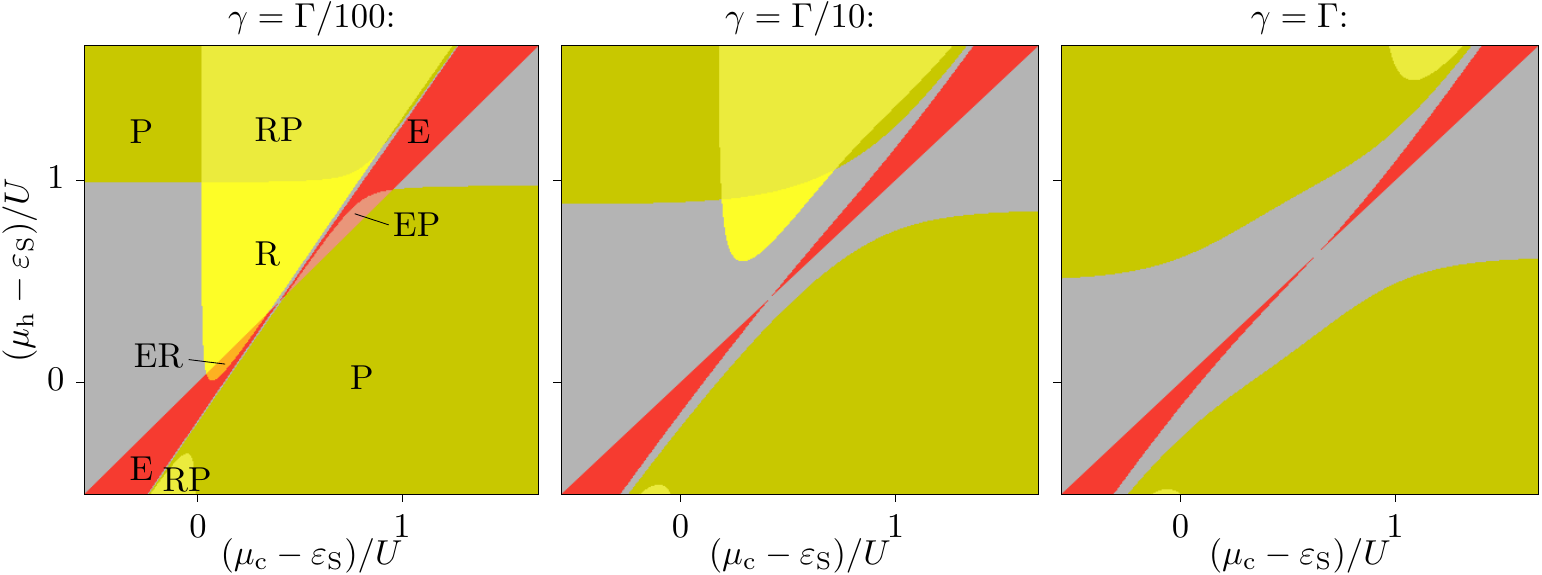}
\caption{\label{fig:mapleak}
Effect of imperfect filtering. The two panels correspond to the same configuration of Fig.~\ref{fig:scheme_regime}, with different $\Gamma_{\rmc 1}=\Gamma_{\rmh 0}=\gamma$. As $\gamma$ increases, more electrons can tunnel through the conductor quantum dot without exchanging energy with the gate. Hence, the purely three-terminal hybrid operations, ER and EP, are more confined and eventually disappear.}
\end{figure*}

The introduction of the generalized free energy currents helps us to interpret the constraints imposed by the second law in a multi-terminal setup. Indeed using definition \eqref{eq:free} the second-law inequality in Eq.~\eqref{eq:second} takes the familiar form $\dot{F}_\mathrm{tot} = \sum_{i} \dot{F}_i  \leq 0$, stating that an overall increase of the free energy is forbidden. 
The role of the reference here is to provide an interpretation of the free energy currents $\dot{F}_i$ as the power that may be retrieved from any terminal $i$ by using reservoir ${\rm r}$ for work extraction purposes (under ideal conditions). In this way, the decrease of $\dot{F}_\mathrm{tot}$ can be interpreted as the overall losses in the power extractable from the hybrid machine, providing an upper bound to the maximum power that the engine may output in any regime.

Now performing the sum over the free energy currents we obtain
\begin{align}
    \dot{F}_\mathrm{tot} &=  - \kB T_{\rm r} \sum_i \dot{S}_i = - \kB T_{\rm r} \dot{S}_\mathrm{tot},
\end{align}
recovering Eq.~\eqref{eq:secondlawF}. Importantly, this implies that the quantification of the power losses in the device through free energy only depends on a single parameter, the temperature of the reservoir that is used as a reference, $T_{\rm r}$. Finally, we remark that while the characterization of the generalized free energy can be done using a generic reference reservoir, it is natural to choose the reference among the reservoirs already present in the setup.

\section{Experimental limitations}

In Sec.~\ref{sec:setup} we assumed a particular configuration for which the conductor dot 
couples to a different reservoir (and only to one), depending on the occupation of the gate dot. In order to achieve this, we considered negligible tunneling rates $\Gamma_{\rmc1}=\Gamma_{\rmh0}=0$. This assumption simplified the discussion considerably by conditioning the electronic transport to transitions that involve the three terminals. This way, two-terminal operations were excluded. 

To achieve this configuration, a proper energy filtering is required. One possibility is provided using superconductors for the cold and hot terminals, choosing $\mu_\rmc>\mu_\rmh$, such that $\varepsilon_\rmS$ lies below the gap of the cold terminal and in the one of the hot terminal, while $\varepsilon_\rmS+U$ is above the gap of the hot terminal and in the one of the cold terminal. Unfortunately, for some cases we will require $\mu_\rmc<\mu_\rmh$ (see e.g. Sec.~\ref{sec:reverse}). An alternative is to consider a triple quantum dot array whose outermost dots act as filters (the left one at $\varepsilon_\rmS$ and the right one at $\varepsilon_\rmS+U$) which are not coupled to the gate~\cite{
sanchez:2011,strasberg:2018}.
The right dot is not necessary if the hot terminal is a superconductor, as described above. 

Experimentally however, totally filtering these transitions might not be perfect e.g., due to higher order tunneling~\cite{dare:2017,walldorf:2017,dare:2019} or finite bandwidth filters~\cite{strasberg:2018}. Here we take into account leaking processes by assuming $\Gamma_{\rmc1}=\Gamma_{\rmh0}=\gamma$. The effect of this coupling on the configuration map is shown in Fig.~\ref{fig:mapleak}. For finite $\gamma$, electrons are allowed to flow along the two conductor terminals without exchanging energy with the gate. This introduces two-terminal processes in the heat exchange between the hot and the cold reservoirs that break the tight-coupling of the currents. The operation boundaries established by Eqs.~\eqref{eq:tight} are hence modified and depend on the different rates. 

As purely three-terminal processes, the hybrid operations ER and EP are affected and reduce their configuration space for low $\gamma/\Gamma$. Increasing the coupling, these are not possible any more, see central panel in Fig.~\ref{fig:mapleak}. In the case when all tunneling rates are equal in the conductor, $\gamma=\Gamma$, only two-terminal like operations are present: E, P and RP. Note that, as in the two-terminal case, refrigeration only comes along with pumping. Furthermore, this occurs out of the region $0<\mu_{\rmc,\rmh}-\varepsilon_\rmS<U$, i.e., where the coupling to the gate becomes irrelevant. Then, we recover the expected behaviour of a two-terminal quantum dot, discussed in App.~\ref{sec:2term}.

In Fig.~\ref{fig:currrevref} we show how currents and efficiencies for a particular configuration are affected by $\gamma$.

\section{Two terminal case}
\label{sec:2term}

\begin{figure}[bth]
\includegraphics[width=0.65\linewidth]{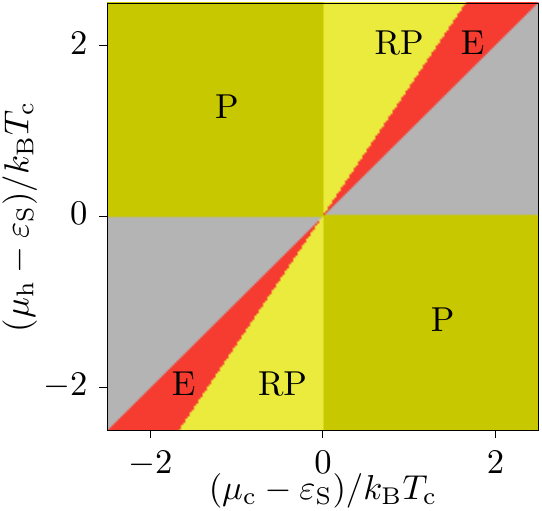}
\caption{\label{fig:mapqd}
Map of the operations of a single-level two terminal quantum dot. It can be obtained as the $U=0$ limiting case of the coupled dot system.
}
\end{figure}

Decoupling the gate dot, the system reduces to a two-terminal single-level quantum dot. For this configuration, all transitions occur at the same energy $\varepsilon_\rmS$, so currents are tightly-coupled~\cite{esposito_thermoelectric_2009}, with 
$\dot{Q}_{\rm c}=(\varepsilon_\rmS-\mu_{\rm c})\dot{N}$, 
$\dot{Q}_{\rm h }=-(\varepsilon_\rmS-\mu_{\rm h})\dot{N}$, and $\dot{W}=-\Delta\mu\dot{N}$, being the particle current
\beq
\dot{N}=\frac{\Gamma_{\rm c}\Gamma_{\rm h}}{\Gamma_{\rm c}+\Gamma_{\rm h}}[f_{\rm c}(\varepsilon_\rmS)-f_{\rm h}(\varepsilon_\rmS)].
\eeq
As we show in Fig.~\ref{fig:mapqd}, this model allows for E, P and RP operations. Cooling is always accompanied by heat pumping for voltages that make the heat engine work in reverse. The opposite is not true: Joule dissipation heats both reservoirs when the dot level lies between the chemical potential of the two leads (so the direction of the particle current flow is dictated by the voltage bias), i.e., when $\mu_{\rm h}{-}\varepsilon_\rmS$ and $\mu_{\rm c}{-}\varepsilon_\rmS$ have opposite signs.

\bibliography{biblio.bib}

\end{document}